\documentclass[aps,prl,floatfix,twocolumn,preprintnumbers,amsmath,amssymb,groupedaddress,showpacs,showkeys]{revtex4-1}

\usepackage{graphicx} 
\usepackage{dcolumn} 
\usepackage{bm} 
\usepackage{color} 
\usepackage{amsmath,amssymb}
\usepackage{mathrsfs}
\usepackage{upgreek}

\usepackage{pdfpages}
\makeatletter
\AtBeginDocument{\let\LS@rot\@undefined}
\makeatother

\begin{document}

\title{Breakdown of the Hebel-Slichter effect in superconducting graphene due to \\ 
        the emergence of Yu-Shiba-Rusinov states at magnetic resonant scatterers}

\author{Denis Kochan}%
	\email[Corresponding author: ]{denis.kochan@ur.de}
 	\affiliation{Institute for Theoretical Physics, University of Regensburg, 93040 Regensburg, Germany}

\author{Michael Barth}%
 	\affiliation{Institute for Theoretical Physics, University of Regensburg, 93040 Regensburg, Germany}
 	
\author{Andreas Costa}%
 	\affiliation{Institute for Theoretical Physics, University of Regensburg, 93040 Regensburg, Germany}
 	
\author{Klaus Richter}%
 	\affiliation{Institute for Theoretical Physics, University of Regensburg, 93040 Regensburg, Germany}
 	
\author{Jaroslav Fabian}%
 	\affiliation{Institute for Theoretical Physics, University of Regensburg, 93040 Regensburg, Germany}


\begin{abstract}
    Employing analytical methods and quantum transport simulations we investigate the relaxation of quasiparticle~spins in graphene proximitized by an $s$-wave superconductor in the 
    presence of resonant magnetic and spin-orbit~active impurities. 
    \emph{Off~resonance}, the relaxation \emph{increases} with \emph{decreasing} temperature when electrons scatter off magnetic impurities---the  Hebel-Slichter~effect---and \emph{decreases} when impurities have spin-orbit coupling. This distinct temperature~dependence (not present in the normal state) uniquely discriminates between the two scattering mechanisms. 
    However, we show that the Hebel-Slichter~picture breaks down \emph{at resonances}. The emergence of Yu-Shiba-Rusinov
    bound states within the superconducting gap redistributes the spectral weight away from magnetic resonances. The result is opposite to the Hebel-Slichter expectation:  the spin relaxation \emph{decreases} with  \emph{decreasing} temperature.  
    Our findings hold for generic $s$-wave superconductors with resonant magnetic impurities, but also, as we show, for resonant magnetic Josephson~junctions.
\end{abstract}

\date{\today}
\maketitle

\paragraph*{Introduction.}
    Superconducting spintronics investigates the interplay between the 
    electron spin phenomena~\cite{Zutic2004} and
    macroscopic quantum coherence of superconducting structures~\cite{EschrigPhysToday2011,EschrigRepProgPhys2015,Linder&Robinson_NP2015}. 
    A versatile platform for superconducting spintronics is offered by  2D layered materials. Indeed, there is a growing family of 2D superconductors---twisted bilayer graphene \cite{Cao-Herrero-MagicAngleBLG_2018,Yankowitz2019}, 2D topological insulators \cite{Sajadi2018,Fatemi2018}, or transition-metal~dichalcogenides~\cite{Shi2015_IonicGatedTMDC-SC,Jo2015_WS2_SC,Xi2016_NbS2-SC,Navarro-Moratalla2016_TaS2-SC,Costanzo2016_MoS2-SC}---which could serve as a source of Cooper pairs. At the same time there are high-mobility 2D (semi)metals and semiconductors whose spin properties, in particular spin relaxation (SR), can profoundly change when proximitized by superconductors. 
    
    Measurements of SR in graphene have not yielded a unique mechanism for electron spin flips~\cite{Tombros2007,Ohishi2007,Pi2010,Yang2011,Avsar2011,Lundeberg2013,Guimaraes2014,Raes2016,Droegeler2016,Dushenko2016}. Ensuing
    intense scientific discussions \cite{Seneor2012,HanKawakamiGmitraFabian_2014,Roche2014,Roche_SpintronicsPerspective_2015,Feng2017,GarciaRoche_ChemSocRev_2018,Rybkin2018} have focused on spin-orbit and exchange impurities as possible 
    culprits. 
    The principal difficulty in setting one against the other lies, unlike in conventional materials \cite{Zutic2004}, in the absence of a systematic temperature behavior of the measured spin relaxation. However, the absence of SR anisotropy \cite{Raes2016} points towards magnetic resonant impurities \cite{Kochan2014_PRL-SR-Graphene,Kochan2015_PRL-SR-BLG} as the main source of spin-flip scattering in graphene. 
    
    Here we show that in (proximitized) superconducting  graphene (SCG) the two types of impurities yield distinct temperature characteristics due to coherence effects. Particularly striking is the prediction that resonant magnetic scatterers cause SR whose temperature dependence is opposite to that predicted by the perturbative Hebel-Slichter effect~\cite{Hebel1957,Hebel1959,Hebel1959-Theory}. Since the non-perturbative analytic and quantum transport simulation methods we use are not specific to graphene, \emph{this prediction applies to resonant scattering in all s-wave superconductors}. Furthermore, we demonstrate that it also applies to superconducting resonant Josephson junctions with magnetic tunnel barriers. 
    
    Although measurements of SR in SCG have not yet been performed (which makes
    theoretical predictions particularly motivating), they are within the current experimental reach. Indeed, proximity induced superconductivity in graphene has been experimentally demonstrated in lateral Josephson junctions \cite{Heersche2007a, Komatsu2012a,Calado2015,Delagrange_PRL_2018}  and vertical stack geometries~\cite{Tonnoir2013,DiBernardo2017_p-wave-SCG}, as well as in 
    alkaline-intercalated graphite~\cite{Li2013,Ludbrook2015,Chapman2016}. The induced superconducting gap ranges from tens of $\upmu\mathrm{eV}$~\cite{Heersche2007a} to 1\,meV~\cite{Rickhaus2012}~($T_{\rm c}\simeq 6.593$\,K). Both $s$-wave~\cite{Heersche2007a} and $p$-wave~\cite{DiBernardo2017_p-wave-SCG} pairings were convincingly demonstrated; see the comprehensive review~\cite{Lee&Lee2018} for more details.

    \begin{figure}
        \begin{center}
        \includegraphics[width=0.80\columnwidth]{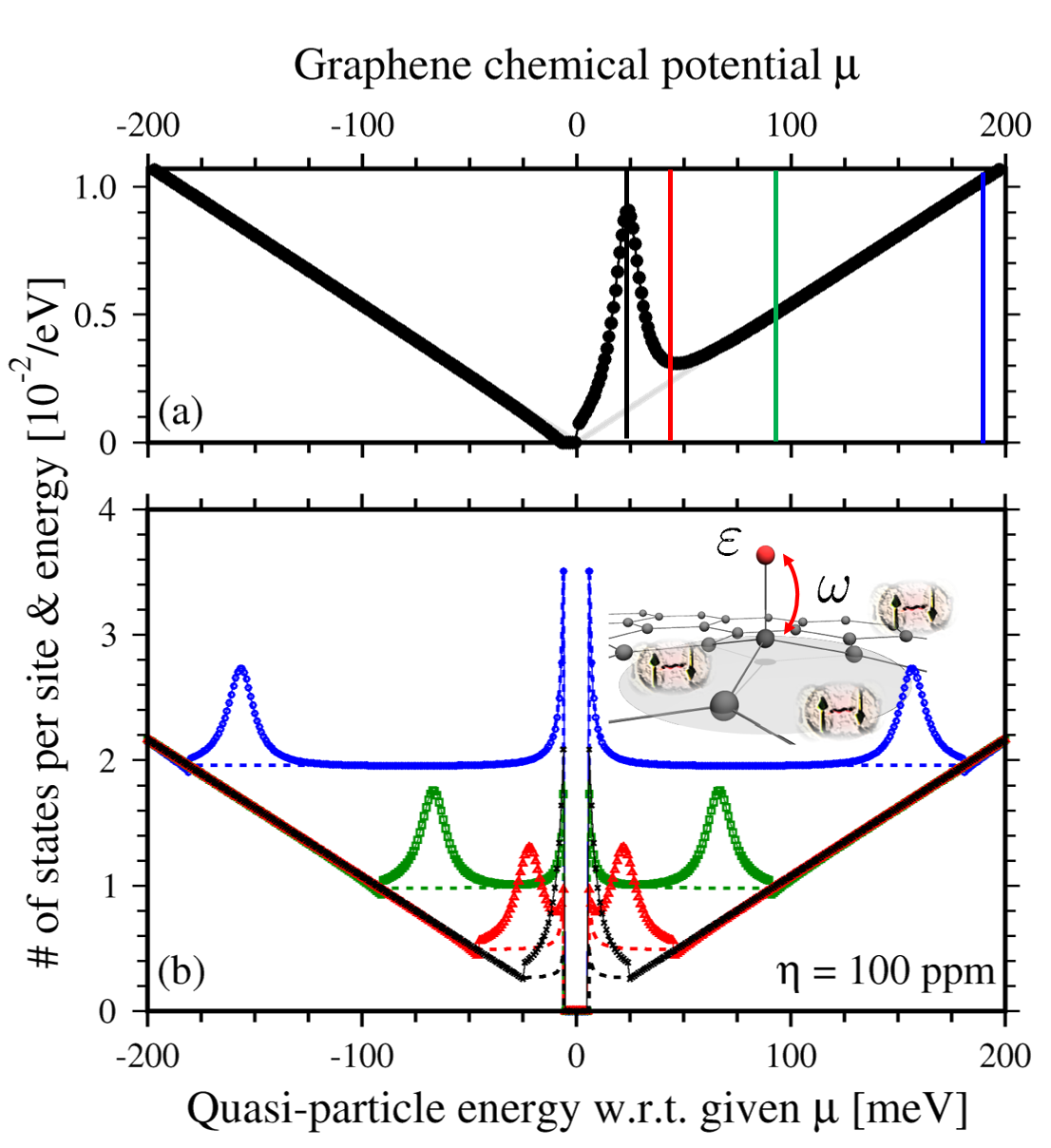}
        \end{center}
        \caption{Calculated density of states. (a) Graphene's normal-state DOS~(black dots) as a function of the chemical~potential~$\mu$ for 100\,ppm of resonant nonmagnetic impurities. A pronounced resonant peak emerges at $\mu=24$\,meV; the gray line is for pristine graphene.
            Black, red, green, and blue vertical lines represent  chemical~potentials
            at which superconductivity is turned on with gap~$\Delta_0=5$\,meV. The corresponding quasiparticle (QP)DOS is shown in~(b). Black symbols stand for $\mu=24$\,meV, red triangles for $\mu=45$\,meV, green squares for $\mu=90$\,meV, and blue circles for $\mu=180$\,meV.
            Dashed lines with the same colors serve as guides for eyes and display the corresponding QPDOS in clean SCG. Resonant  enhancements near the coherence~peaks appear for chemical~potentials close to the normal-phase~resonances.
            \emph{Inset:}~Tight-binding description of an adatom absorbed on SCG. For DOS we used hybridization~$\omega=5.5$\,eV and on-site~energy~$\varepsilon=0.26$\,eV.}
            \label{Fig:1}
        \end{figure}
        \begin{figure*}
            \begin{center}
              \includegraphics[width=1.9\columnwidth]{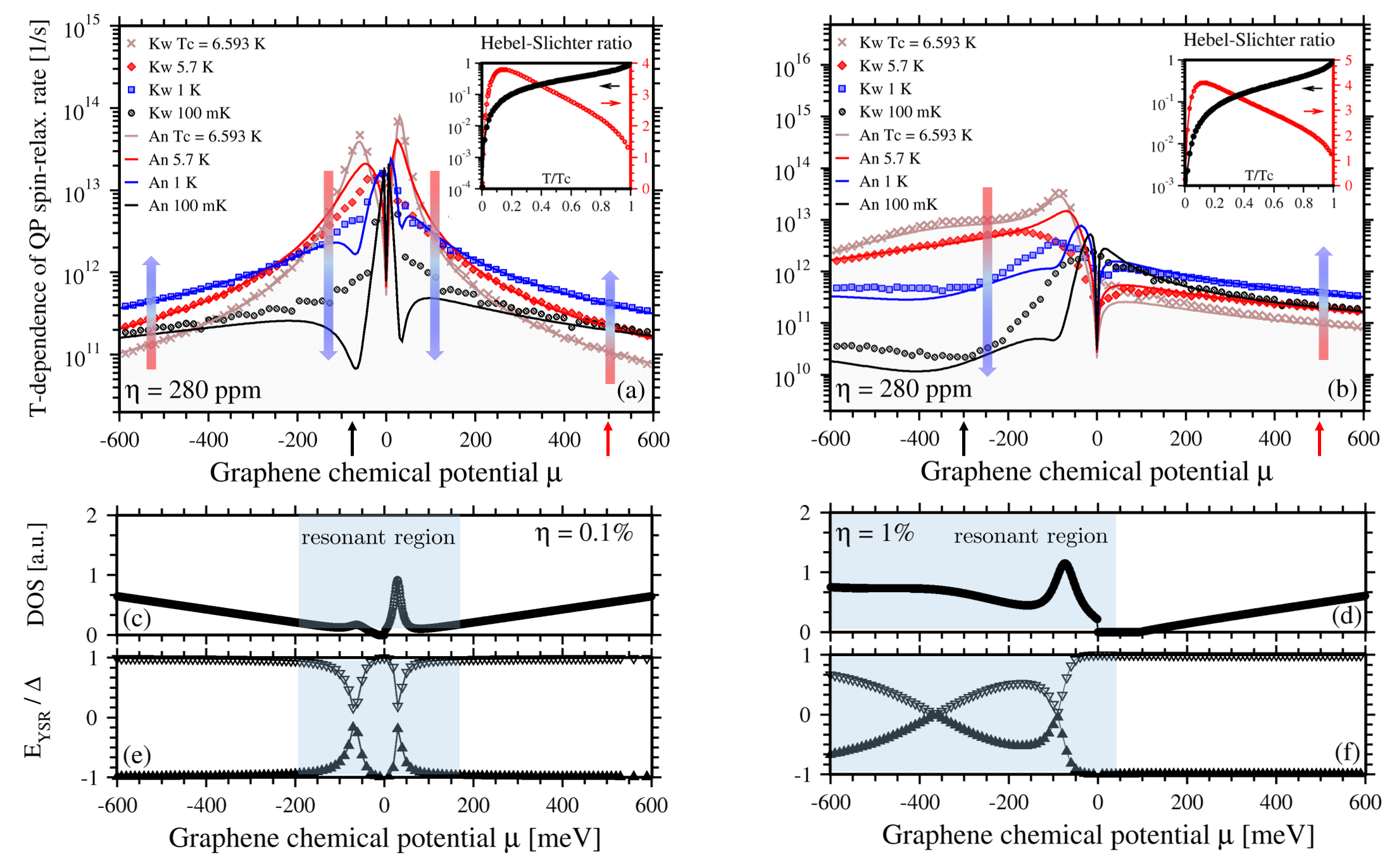}
            \end{center}
            \caption{Quasiparticle SR~rates in SCG~(as functions of~$\mu$) at different temperatures~(different colors) for $280$\,ppm of (a)~hydrogen and (b)~fluorine magnetic~impurities, solid lines---analytical calculations, symbols---\emph{Kwant} simulations. 
            The SR~rates  \emph{increase off resonance} whereas they \emph{decrease in the resonances}. 
            Rainbow arrows~(coded in colors of temperature descent) indicate the SR~rates' increasing or decreasing trends with lowered~$T$ compared to the normal phase.
            The insets show the corresponding Hebel-Slichter~ratios, $(1/\tau_s^{\rm SC})\bigl/(1/\tau_s^{\rm N})$, as functions of $T/T_{\rm c}$ and at two representative Fermi~energies~(indicated by black and red arrow ticks on the horizontal axis): \emph{resonant}~($\mu=-80$\,meV for hydrogen and $\mu=-300$\,meV for fluorine; see black circled data with values at left \emph{logarithmic} axis) and \emph{off~resonant}~($\mu=500$\,meV for both cases; see red circled data with values at right \emph{linear} axis). \emph{Kwant} simulation used a scattering region with $W=299\,a$ and $L=5\,a$, a small misalignment around charge neutrality originates from finite-size effects. 
            Panels~(c)~and~(d)~show the normal-phase~DOS obtained from analytical calculation in the presence of magnetic moments, and \emph{resonant}~(shaded) and \emph{off-resonant}~(white) doping regions; to increase the visibility, the impurity~concentrations were exaggerated.
            Panels~(e)~and~(f)~obtained form the poles of the analytical Green functions display the subgap YSR~states' energies for hydrogen and fluorine~(as functions of~$\mu$). Smaller SR~rates in (a)~and~(b)~are correlated with the normal-phase~resonances in~(c)~and~(d), and the YSR~states in~(e)~and~(f)~with energies deep inside the gap.}
            \label{Fig:2}
        \end{figure*}
    \begin{figure}
        \begin{center}
        \includegraphics[width=0.99\columnwidth]{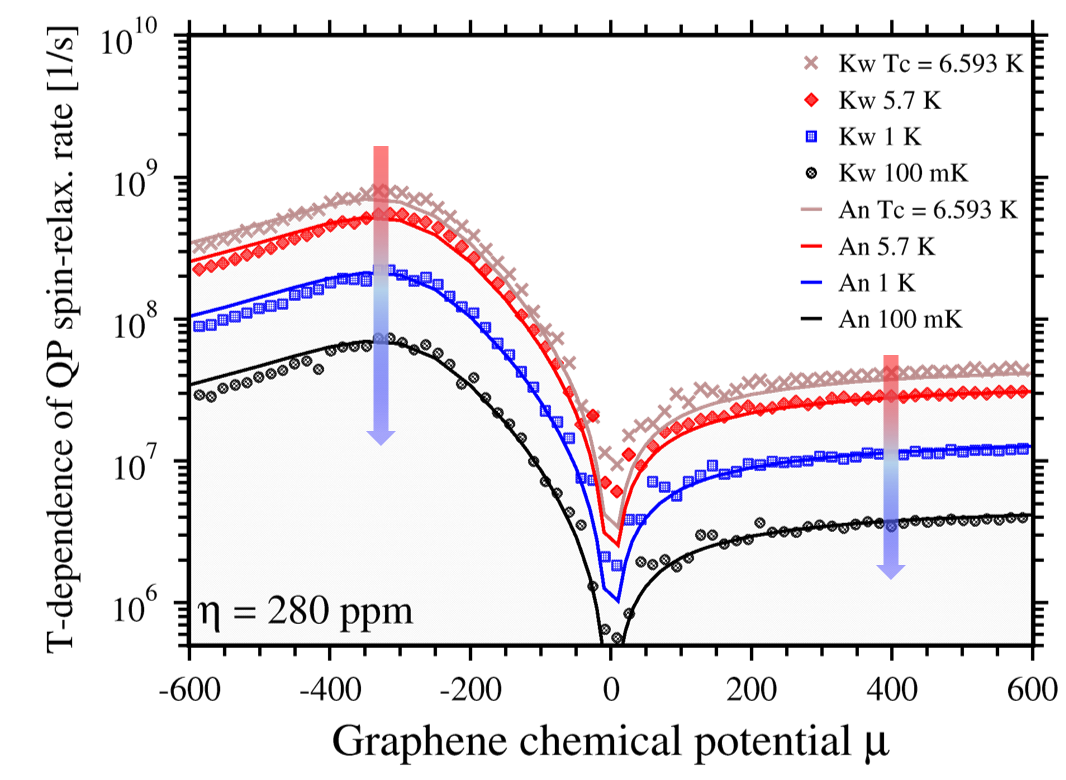}
        \end{center}
        \caption{SR~rates in SCG~(as functions of~$\mu$) at different temperatures~(different colors) due to locally enhanced 
             SOC for $280$\,ppm of fluorine impurities, solid lines---analytical calculations, symbols---\emph{Kwant} simulations. The SR~rates decrease almost uniformly with lowering 
             temperature~$ T $; their decrease becomes steeper and would saturate as $T\rightarrow 0$. Similarly to the normal 
             state, the SR~rates are enhanced at resonances. Rainbow arrows ~(coded in colors of temperature descent) indicate the SR~rates' decreasing trend with lowered $T$.}
        \label{Fig:3}
        \end{figure}
   

\paragraph*{Rationale.}
Yafet showed ~\cite{Yafet1983}, using the first order perturbation theory, that the SR~rate in superconductors is modified from the normal state as $1/\tau_s^{\rm SC}\sim\langle(u_{\mathbf{k}}u_{\mathbf{q}}\pm v_{\mathbf{k}}v_{\mathbf{q}})^2/\tau_s^{\rm N}\rangle$, where
$u$ and $v$ are standard BCS~coherence~factors and $\langle\cdots\rangle$ denotes thermal broadening. The plus sign applies to interactions that are  odd under time reversal, such as exchange. As a result, $1/\tau_s^{\rm SC}>1/\tau_s^{\rm N}$, which is the Hebel-Slichter~effect~\cite{Hebel1957,Hebel1959,Hebel1959-Theory}. On the other hand, the minus sign is for time-reversal symmetric interactions such as spin-orbit coupling, in which case  
$1/\tau_s^{\rm SC}<1/\tau_s^{\rm N}$, as
demonstrated experimentally for aluminum~\cite{Yang_Parkin2010,Quay_Aprili_Strunk_2015}
\footnote{For more details about the charge and spin~accumulation of QPs in superconductors, their non-equilibrium separation, and relaxation, consult Refs.~\cite{Zhao1995,Yamashita_Maekawa2002,Quay2013}.}. 
We will see that in SCG $1/\tau_s^{\rm N}$ and $1/\tau_s^{\rm SC}$ can differ
by orders of magnitude due to the coherence effects, which allows 
for an unprecedented experimental feasibility \emph{to disentangle the dominant SR~mechanism}.
Our methodology overcomes two shortcomings
of the standard theory of Yafet, Hebel, and Slichter. First, we calculate the SR rate to all orders of perturbation theory allowing us to consider resonant scattering and, second, we also include
sub-gap Yu-Shiba-Rusinov~(YSR)~states~\cite{Yu1965,Shiba1968,Rusinov1968,*Rusinov1968alt, CostaPRB2018} which have no normal-state
counterpart and which take away considerable spectral weight from the scattering states.

\paragraph*{Model~and~Methodology.}
    To describe SCG we use the  tight-binding~Hamiltonian~\cite{Uchoa2007_SC-Graphene},
    \begin{multline}
        H_{g}=-\sum_{m, n, \sigma} (t\delta_{\langle mn\rangle}+\mu\delta_{mn})c^\dagger_{m\sigma} c^{\phantom{\dagger}}_{n\sigma} \\
        +\Delta\sum_{m} c^\dagger_{m\uparrow} c^\dagger_{m\downarrow}+\rm{h.c.} ;
    \end{multline}
    $t=2.6$\,eV stands for the conventional nearest~neighbor~hopping, $\mu$ for the chemical~potential~(doping~level) taking the normal-phase's charge neutrality~point as a reference, and $\Delta$ models the $T$-dependent global on-site $s$-wave~pairing. We assume the BCS-like temperature~dependence~$\Delta(T)=\Delta_0 \tanh{[1.74 \sqrt{T_{\rm c}/T - 1}]}$ with realistic proximity values of $\Delta_0=1$\,meV and $T_\mathrm{c}\simeq 6.593
    $\,K.
    The operator $c_{m\sigma}^{(\dagger)}$ annihilates~(creates) an electron with spin $\sigma$ at the graphene lattice~site~$m$, $\delta_{mn}$ represents the usual Kronecker~symbol, and $\delta_{\langle mn\rangle}$ its nearest-neighbor~analog~(unity for the direct graphene nearest-neighbor sites and zero otherwise). 
    Orbital interactions with an adatom---annihilation and creation operators $d_{\sigma}$ and $d_{\sigma}^{\dagger}$---are governed by the impurity-site hybridization~$\omega$, on-site energy~$\varepsilon$, and the proximity pairing~$\Delta$, combining into~\cite{Ratto1967} (see the inset of Fig.~\ref{Fig:1})
    \begin{equation}
        V_o=\sum_{\sigma} [(\varepsilon-\mu) d^\dagger_\sigma d^{\phantom{\dagger}}_\sigma +\omega d^\dagger_\sigma c^{\phantom{\dagger}}_{0\sigma}]+\Delta d^\dagger_\uparrow d^\dagger_\downarrow +\rm{h.c.} ;
    \end{equation}
    This orbital~perturbation is complemented by a local spin-dependent~term $V_s$ comprising (1) exchange interaction, $V_s^{(1)}=-J\,\mathbf{s}\cdot\mathbf{S}$, between the itinerant electron spin~$\mathbf{s}$ and the impurity $ 1/2 $-spin~$\mathbf{S}$~\cite{Kochan2014_PRL-SR-Graphene}, and (2)~local SOC in the vicinity of an adatom~\cite{Gmitra2013_SOC-in-H-Graphene,Irmer2015-SOC-F-Graphene,Zollner2016-SOC-Methyl,FrankPRB2017Copper,Kochan2017-SOC-Model-Ham,SM}~
    To be specific, we use hydrogen and fluorine adatoms, both of which induce sizable SOC~enhancements~\cite{Gmitra2013_SOC-in-H-Graphene,Irmer2015-SOC-F-Graphene} and can also carry magnetic~moments~\cite{Yazyev2010_RepProgPhys,Xie2011,Hong2012,Nair2012,Gonzalez-Herrero2016,Szalowski2016,Susanne_Jun_Jaro_Collab,Sousa2019}. We also assume low concentrations $\eta$~(per carbon atom) of dilute spin-active~impurities~\cite{Zuckermann1965} for the independent scatter picture to be valid. 
    
    Our methodology employs the \emph{full analytical approach}, calculating the spin-flip~rates from the T-matrix, as well as \emph{numerical \textit{Kwant} simulations} of spin-flip scattering probabilities, providing together coherent and consistent qualitative and quantitative pictures. Detailed methodology is in Supplementary Material~\cite{SM}.

    \paragraph*{Results.}
    Adatoms on graphene give rise to resonances~\cite{Skrypnyk2007,Wehling2010_ResonantScattering,Ferreira2015_ZeroMode,IrmerPRB2018TopBridgeHollow}, which strongly modify graphen's transport properties~\cite{Skrypnyk2010,Ferreira2014_ExtrinsicSHE_ResonantScattering,Stabile2015,Garcia2016_KuboBastin,KatochPRL2018,IrmerPRB2018TopBridgeHollow,JoengsuLeePRB2019LandauLevels},  
    particularly when they lie close to the Dirac~point.
    Figure~\ref{Fig:1} demonstrates how normal-state~resonances affect the population of quasiparticle states in SCG at different chemical potentials. Panel~\ref{Fig:1}(a)~shows the DOS of graphene covered by 100\,ppm of resonant \emph{nonmagnetic}~impurities, while panel~\ref{Fig:1}(b) displays the corresponding quasiparticle DOS (QPDOS). We present resonant and off-resonant doping~limits and conclude that  QPDOS gets strongly modified near the coherence~peaks as~$\mu$ approaches the normal-state~resonance. This makes sense since 
    BCS theory gives ~QPDOS$(E)=[E/\sqrt{E^2-\Delta^2}]$ DOS$(\mu)$.

    Figure~\ref{Fig:2} illustrates various characteristics of spin-flip~scattering off \emph{magnetic}~impurities in normal and superconducting graphene for two representative impurities: hydrogen---panels~\ref{Fig:2}(a),(c),(e)---and fluorine---panels~\ref{Fig:2}(b),(d),(f).
    Particularly, Figs.~\ref{Fig:2}(a)~and~(b)~display the SR~rates in SCG for different temperatures in the presence of 280\,ppm of magnetic~impurities,  
    as a function of the chemical~potential~$\mu$~and the superconducting~gap with temperature 
    $T$. Solid lines corresponds to analytical T-matrix calculation, and the symbols with the same color to the corresponding \emph{Kwant} simulation, for details see~\cite{SM}.
    Hydrogen~\cite{Kochan2014_PRL-SR-Graphene} with a nonzero magnetic moment---$\omega=7.5$\,eV, $\varepsilon=0.16$\,eV, and $J=-0.4$\,eV---gives rise to a narrow normal-state resonant region near the Dirac~point; see the corresponding magnetic DOS in~Fig~\ref{Fig:2}(c)~(concentration $\eta=0.1\%$ is exaggerated for resolution purposes). In contrast, fluorine---$\omega=5.5$\,eV, $\varepsilon=-2.2$\,eV, and $J=0.5$\,eV---develops~\cite{Irmer2015-SOC-F-Graphene,Susanne_Jun_Jaro_Collab} a wide resonance~region spreading below the Dirac~point; see the magnetic DOS in~Fig~\ref{Fig:2}(d)~(again with the elevated concentration $\eta=1\%)$. 
    
    The striking impact of resonances on quasiparticle SR~rates is seen from Figs.~\ref{Fig:2}(a)~and~(b). The shaded regions show the SR~rate in the normal~phase~($T=T_\mathrm{c}$). Lowering $T$ below superconducting $T_c$ we observe an intriguing behavior: for off-resonant doping~regions, $1/\tau_s^{\rm SC}>1/\tau_s^{\rm N}$ in accordance with the Hebel-Slichter scenario~\cite{Hebel1957,Hebel1959,Hebel1959-Theory}, whereas quasiparticle SR sharply drops, ~$1/\tau_s^{\rm SC}\ll 1/\tau_s^{\rm N}$, at resonances. 
    To further quantify those effects, the insets of Figs.~\ref{Fig:2}(a)~and~(b) represent the corresponding \emph{Hebel-Slichter~ratios}, $(1/\tau_s^{\rm SC})\bigl/(1/\tau_s^{\rm N})$, taken at two representative $\mu$'s as functions of~$T/T_{\rm c}$. For the off-resonant value of~$\mu=500$\,meV, both impurity~cases lead to a notable enhancement of the superconducting SR~rates by almost a factor of four~(graphs with red symbols), while we see a strong decrease of the SR~rates by almost three orders of magnitude~(graphs with black symbols in logarithmic scale) inside the resonant regions---we use $\mu=-80$\,meV for hydrogen and~$\mu=-300$\,meV for fluorine. 
    This can serve as a powerful experimental evidence---observing strongly depleted SR~rates in the SCP when lowering $T$ signifies the presence of \emph{resonant} magnetic impurities.

    To explain this peculiar resonant depletion of QP~SRs in SCG, which is at odds with the well-understood resonant enhancement~\cite{Kochan2014_PRL-SR-Graphene,Kochan2015_PRL-SR-BLG,Soriano2015,Wilhem2015:PRB,MIRANDA2019,Thomsen2015} in the normal~phase, we calculate the energies~(poles of T-matrix, \cite{Loktev2015}) of the corresponding YSR~states~\cite{Yu1965,Shiba1968,Rusinov1968,*Rusinov1968alt} which develop around magnetic impurities~\cite{Wehling2008,Lado2016a}, see Figs.~\ref{Fig:2}(e)~and~(f). 
    We deduce that the YSR~states lie deep inside the superconducting~gap at doping levels that correspond to resonances in the normal-phase of graphene. This offers an explanation why the SR~rates of quasiparticles sharply decrease. 
    Resonant scattering of QPs counts many contributions, especially those from multiple~scatterings and those from virtual states' tunnelings. 
    Schematically, the scattering matrix element can be written as $V_{\alpha\alpha}+V_{\alpha\mathrm{I}}\tfrac{|\mathrm{I}\rangle\,\langle \mathrm{I}|}{E_\alpha-E_\mathrm{I}+i0_{+}}V_{\mathrm{I}\alpha}+\ldots$, where $E_\mathrm{I}$ represents the energy of any intermediate state~(extended or subgap), and $E_\alpha$ stands for the energy of an incident QP~state. The dominant elements~$V_{\alpha\mathrm{I}}$ for spin-flip processes arise from a strong overlap of quasiparticle states $\alpha$ with the magnetic impurity~level~$\mathrm{I}$=YSR since only the latter gives rise to QP spin~flips. 
    While the matrix-elements~$V_{\alpha\mathrm{YSR}}$ are roughly the same for all extended QP states $\alpha$ at the coherence~peaks, independently of doping (impurity in our model acts like $\delta$-function), this is not true for the energy differences~$E_\alpha-E_{\mathrm{YSR}}$ in the denominator. Those are, in~fact, small in the off-resonant region since $E_{\mathrm{YSR}}$ are aligned with the superconducting gap~edges and become large at the resonances, where $E_{\mathrm{YSR}}$ are deep inside the gap. 
    That this must cause the reduced SR is clear from the $T$-dependence of the SR~rates; for higher $T$, $\Delta$ gets smaller and hence also the difference~$E_\alpha-E_{\mathrm{YSR}}$. 
    The opposite conclusion would follow from Yafet's~formula, which does not account for the formation of subgap bound~states. 
    
    At very low temperatures---data for $T=100$\,mK displayed by dashed lines in Figs.~\ref{Fig:2}(a)~and~(b)---, the SR~rates drop down at low dopings since the related QPDOS, $\frac{E}{\sqrt{E^2-\Delta^2}}\,\text{DOS}(\mu)(-\tfrac{\partial g}{\partial E})$, 
    becomes substantially suppressed upon thermal 
    smearing. However, at larger dopings this suppression is counteracted by the elevated~$\text{DOS}(\mu)$.

    Finally, we turn to SR due to spin-orbit active impurities. Figure~\ref{Fig:3} displays the calculated SR~rates in SCG as functions of doping at different temperatures for fluorine~impurities, see~\cite{SM} for the hydrogen which shows the same qualitative features. 
    We find that the SR~rates in SCG decrease (as predicted by Yafet~\cite{Yafet1983}) by one order of magnitude with lowering $T$, giving sizable experimental signals. 
    At resonances SR rates get enhanced, similarly to 
    what was predicted for the normal state~\cite{Bundesmann_SR_SOC}. 
    This is because a quasiparticle in a resonant state has sufficient time to experience SOC and flip its spin, what gives rise to the increased SR rates, although somewhat less than in the normal state due to lower group~velocities of quasiparticles. 
    Thus, observing a global decrease of the SR~rate with lowered~$ T $ spanning a wide doping region would be an unprecedented experimental signal for SOC-dominated SR. 
    
    Although motivated by superconducting spintronics we used primarily SCG to perform detailed simulations, our findings are also qualitative and thus expected to be valid for \emph{resonant scattering off magnetic impurities in generic s-wave superconductors}. In fact, we have analyzed a simple 1D model of a resonant Josephson junction with magnetic tunnel barriers \cite{SM}, which explicitly demonstrates the reduction of spin-flip probabilities of quasiparticles at resonances due to the appearance of deep immersed YSR states.

    \paragraph*{Conclusions.}
    We have shown that the spin relaxation of quasiparticles due to resonant magnetic impurities 
    has the opposite temperature dependence from what is predicted as the Hebel-Slichter effect. The reason is emergence of sub-gap YSR states which redistribute the spectral weight away from the resonances. The anomalous decrease of resonant SR can span three-to-four orders of magnitude, making it a robust and verifiable experimental tool.


    \paragraph*{Acknowledgements.}
    D.K.~thanks Drs.~Ferenc~Simon, Yuriy Pogorelov, Lucia~Komendov\'{a}, and Benedikt~Scharf for useful discussions, and Dr.~Jeongsu~Lee for helpful tips regarding numerical implementation. 
    The authors acknowledge support from Deutsche~Forschungsgemeinschaft~(DFG, German~Research~Foundation)---Project-ID~314695032---SFB~1277~(Subprojects~A07, A09, B07), the EU~Seventh~Framework~Programme under Grant~Agreement~No.~604391~(Graphene~Flagship), and the International Doctorate~Program Topological~Insulators of the Elite~Network of Bavaria.

\bibliography{library}

	\onecolumngrid
	\newpage
	

\section*{Supplementary material}

# Breakdown of the Hebel-Slichter effect in superconducting graphene due to the emergence of Yu-Shiba-Rusinov states at magnetic resonant scatterers

Denis Kochan, Michael Barth, Andreas Costa, Klaus Richter, and Jaroslav Fabian
*Institute for Theoretical Physics, University of Regensburg, 93040 Regensburg, Germany*

The following Supplemental Material presents in detail spin-relaxation fundamentals in connection to analytical T-matrix approach, and, primarily, to its superconducting *Kwant-package implementation*. The last section discusses the break-down of the Hebel-Slichter effect in resonant Josephson junctions, what provides an evidence that the effect is more universal and extends beyond superconducting graphene.

Spin relaxation of quasiparticles (QPs) in the superconducting phase depends on the underlying scattering mechanism, namely its time-reversal parity [S1]. The latter determines how the electron and hole transition amplitudes combine giving rise to the final spin-flip rate. As pointed out by Yafet [S2], the spin relaxation rate in the superconducting phase, $1/\tau_s^{\rm SC}$, relates—within first-order perturbation theory—to its normal-phase counterpart, $1/\tau_s^{\rm N}$, by

$$1/\tau_s^{\rm SC} \sim \langle (u_{\mathbf{k}} u_{\mathbf{q}} \pm v_{\mathbf{k}} v_{\mathbf{q}})^2 / \tau_s^{\rm N} \rangle. \quad (\text{S1})$$

Here, $u$ and $v$ are the standard BCS coherence factors of the QP wave functions participating in scattering, and $\langle \cdots \rangle$ represents thermal broadening over the QP energies. Consequently, the spin relaxation rate in the superconducting phase increases or decreases depending on the relative sign between the coherence factors. The plus (minus) sign applies to perturbations that are odd (even) w.r.t. time-reversal symmetry, e.g., magnetic impurities (local SOC fields), eventually giving rise to a larger (smaller) $1/\tau_s^{\rm SC}$ compared to $1/\tau_s^{\rm N}$. As demonstrated in the main text for superconducting graphene, those differences vary with the chemical potential and temperature, and can even change by a few orders of magnitude. Thus conducting the same spin relaxation measurement in the normal and superconducting phase of graphene, and comparing $1/\tau_s^{\rm N}$ to $1/\tau_s^{\rm SC}$ would provide an unprecedented experimental feasibility *to disentangle the dominant spin relaxation mechanism.*

The qualitative physical understanding is rather intuitive. On the one hand, QPs have well-defined spins and almost the same mass as normal-phase carriers, but energy dependent effective charges $q = (u^2 - v^2)\,e_{\rm el.}$, especially for QPs in the coherence peaks ($u^2 \simeq v^2$) that dominate scattering at $T < T_{\rm c}$. Therefore, 'direct charge-charge' interaction is less pronounced (superconducting pairing enfeebles Coulomb repulsion) and the 'indirect spin-spin exchange' takes over, giving $1/\tau_s^{\rm SC} > 1/\tau_s^{\rm N}$. This effect is known as *Hebel-Slichter effect* [S3–S5]. On the other hand, not only the QP charges diminish, but also their group velocities $v_{\rm SC} \simeq |(u^2 - v^2)|\,v_{\rm N}$ and likewise their momenta. Since spin-orbit coupling (SOC) couples QP spins to their momenta

$$\text{SOC}_{\rm SC} \approx m(\mathbf{v}_{\rm SC} \times \mathbf{L})\cdot\mathbf{s} \approx |(u^2 - v^2)| m(\mathbf{v}_{\rm N} \times \mathbf{L})\cdot\mathbf{s} \approx \text{SOC}_{\rm N},$$

the effective SOC strength significantly decreases in the superconducting phase, implying $1/\tau_s^{\rm SC} < 1/\tau_s^{\rm N}$.

## SPIN-ORBIT COUPLING HAMILTONIAN

In the main text, we use the local SOC Hamiltonians $V_s^{(2)}$ that capture spatially localized spin-orbit fields adatoms induce in graphene. Those tight-binding model Hamiltonians are based on relevant local symmetries [S6] and their couplings are fitted to first-principles calculations; see Ref. [S7] for more details regarding hydrogen and Ref. [S8] for fluorine.

Since the SOC of pristine graphene is weak, about $40\,\mu\text{eV}$ [S9], it does not play a significant role, and we disregard that global SOC contribution. The defect region in our model consists of the adatomized carbon (site $m = 0$) and its three nearest (nn), $\text{C}_{\rm nn}$, and six next-nearest (nnn), $\text{C}_{\rm nnn}$, neighbors. A minimal effective SOC Hamiltonian in local atomic basis that is consistent

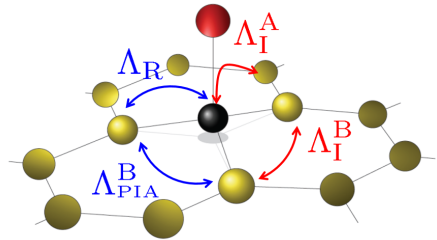

FIG. S1. Schematic representation of the local SOC strengths in the vicinity of the adatom (red), which eventually enter the SOC Hamiltonian $V_s^{(2)}$. Blue and red arrows label spin-flipping and spin-conserving SOC hoppings, connecting specific nearest or next-nearest neighbor carbon atoms (yellow).

with the local symmetries reads then as

$$V_s^{(2)} = \frac{i\Lambda_I^A}{3\sqrt{3}} \sum_{m \in C_{nnn}} \sum_\sigma c_{0\sigma}^\dagger (\hat{s}_z)_{\sigma\sigma} c_{m\sigma} + \text{h.c.}$$
$$+ \frac{i\Lambda_I^B}{3\sqrt{3}} \sum_{\substack{m,n \in C_{nn} \\ m \neq n}} \sum_\sigma c_{m\sigma}^\dagger \nu_{mn} (\hat{s}_z)_{\sigma\sigma} c_{n\sigma}$$
$$+ \frac{2i\Lambda_R}{3} \sum_{m \in C_{nn}} \sum_{\sigma \neq \sigma'} c_{0\sigma}^\dagger (\hat{s} \times \boldsymbol{d}_{0m})_{z,\sigma\sigma'} c_{m\sigma'} + \text{h.c.}$$
$$+ \frac{2i\Lambda_{PIA}^A}{3} \sum_{m \in C_{nnn}} \sum_{\sigma \neq \sigma'} c_{0\sigma}^\dagger (\boldsymbol{d}_{0m} \times \hat{s})_{z,\sigma\sigma'} c_{m\sigma'} + \text{h.c.}$$
$$+ \frac{2i\Lambda_{PIA}^B}{3} \sum_{\substack{m,n \in C_{nn} \\ m \neq n}} \sum_{\sigma \neq \sigma'} c_{m\sigma}^\dagger (\boldsymbol{d}_{mn} \times \hat{s})_{z,\sigma\sigma'} c_{n\sigma'},$$
(S2)

where $\hat{\boldsymbol{s}}$ represents an array of Pauli matrices acting in spin space. The sign factor $\nu_{mn}$ equals $-1$ ($+1$) if the next-nearest neighbor hopping $n \to l \to m$ via a common neighbor $l$ becomes (counter)clockwise oriented and the unit vector $\boldsymbol{d}_{mn} = \frac{\boldsymbol{R}_m - \boldsymbol{R}_n}{|\boldsymbol{R}_m - \boldsymbol{R}_n|}$ points from site $n$ to $m$. The first two terms in Eq. (S2) are the local *intrinsic* SOCs associated with sublattices A and B, respectively, the third is the local *Rashba* SOC, and the last two terms are the local *pseudospin inversion asymmetry (PIA)*-induced SOC terms for sublattices A and B; for more details, see [S6]. The graphical representation of local SOC hoppings is depicted in Fig. S1. The numerical values of these parameters for hydrogenated and fluorinated graphene are summarized in Tab. S1. We adopted those values in our numerical and analytical calculations of the rates in superconducting graphene.

## SPIN RELAXATION RATES IN SUPERCONDUCTING GRAPHENE— ANALYTICS *VS.* NUMERICS

### Analytical computational scheme

The theory of spin relaxation is well developed and a commonly used subject [S10]; for the mechanisms discussed here—Elliott-Yafet and resonant exchange model—we refer the reader to the proceedings reviews [S11] and [S12]. The basic idea is to start from the

TABLE S1. Spin-orbital tight-binding parameters (in meV), entering the model Hamiltonian $V_s^{(2)}$.

| Adatom | $\Lambda_I^A$ | $\Lambda_I^B$ | $\Lambda_R$ | $\Lambda_{PIA}^A$ | $\Lambda_{PIA}^B$ |
|---|---|---|---|---|---|
| Hydrogen | $-0.21$ | 0 | 0.33 | 0 | 0.77 |
| Fluorine | 0 | 3.3 | 11.2 | 0 | -7.3 |

transition rates $W_{\mathbf{k},\uparrow|\mathbf{q},\downarrow}$ and $W_{\mathbf{k},\downarrow|\mathbf{q},\uparrow}$ giving the probability per unit time that an electron with momentum $\mathbf{q}$ scatters into state with momentum $\mathbf{k}$, simultaneously flipping its spin. Those rates are obtained from the microscopic theory—model Hamiltonian of the host system, $H_0$, and perturbation, $V$, that causes spin flips (e.g., SOC, magnetic moments, interaction with phonons, absorption of circularly polarized light, etc.). Transition rates are obtained from Fermi's golden rule

$$W_{\mathbf{k},\sigma|\mathbf{q},\sigma'} = \frac{2\pi}{\hbar} \big|\langle \mathbf{k},\sigma|V|\mathbf{q},\sigma'\rangle\big|^2 \delta(E_{\mathbf{k},\sigma} - E_{\mathbf{q},\sigma'}), \quad \text{(S3)}$$

or the generalized Fermi golden rule that goes beyond first order perturbation theory

$$W_{\mathbf{k},\sigma|\mathbf{q},\sigma'} = \frac{2\pi}{\hbar} \big|\langle \mathbf{k},\sigma|\mathbb{T}|\mathbf{q},\sigma'\rangle\big|^2 \delta(E_{\mathbf{k},\sigma} - E_{\mathbf{q},\sigma'}). \quad \text{(S4)}$$

Here, $\mathbb{T}$ is the so-called T-matrix, which is formally given as

$$\mathbb{T} = V + V\frac{1}{E - H_0 + i\epsilon}V +$$
$$+ V\frac{1}{E - H_0 + i\epsilon}V\frac{1}{E - H_0 + i\epsilon}V + \cdots \quad \text{(S5)}$$
$$= \frac{V}{1 - \frac{1}{E - H_0 + i\epsilon}V}$$

and counts for the multiple scattering processes that are important for the proper account of resonances [S13, S14].

For perturbations $V$ that are spatially local, what is the case considered here, the expression for the T-matrix can be handled explicitly in the local atomic (or Wannier) basis—the resulting object is a finite-size matrix effectively encoding hoppings between different atomic orbitals residing on different lattice sites. The price to be paid is to invert the operator $E - H_0 + i\epsilon$ (giving the retarded Green's resolution $\mathbb{G}_0$), and express its elements in the Wannier basis (basically the Fourier transformation). Exactly those steps were carried out analytically for the Hamiltonian $H_0 = H_g$ describing the unperturbed superconducting graphene, see Eq. (1) in the main text. The resulting formulas for $\mathbb{G}_0 = (E - H_0 + i\epsilon)^{-1}$ in the Wannier basis are too long to be concisely presented here.

Knowing the transition rates $W_{\mathbf{k},\sigma|\mathbf{q},\sigma'}$, one forms the detailed balance equations that govern the time evolution of the spin distribution functions $f_{\mathbf{k},\uparrow}(t) = f_\mathbf{k}^0 + \delta f_{\mathbf{k},\uparrow}(t)$ and $f_{\mathbf{k},\downarrow}(t) = f_\mathbf{k}^0 + \delta f_{\mathbf{k},\downarrow}(t)$. Those are assumed to be not far from local thermodynamic equilibrium—characterized by the chemical potential $\mu$, temperature $T$, and the equilibrium Fermi-Dirac distribution $f_\mathbf{k}^0(E, \mu, T)$ [outside of this section we denote the latter by $g(E)$; moreover, we assume that the equilibrium has no net magnetization].

The spin relaxation time $\tau_s$ (also called $T_1$-time) is

defined from the equation

$$-\frac{\partial}{\partial t}\sum_{\mathbf{k}} f_{\mathbf{k},\uparrow}(t) - f_{\mathbf{k},\downarrow}(t) =: \frac{1}{\tau_s}\sum_{\mathbf{k}} f_{\mathbf{k},\uparrow}(t) - f_{\mathbf{k},\downarrow}(t). \quad \text{(S6)}$$

The derivation from the detailed balance principle in the case of, e.g., Elliott-Yafet spin-relaxation mechanism can be found in Ref. [S11]. The final formula for $1/\tau_s(\mu,T)$ reads

$$\frac{1}{\tau_s} = \sum_{\mathbf{k},\mathbf{q}} W_{\mathbf{k},\uparrow|\mathbf{q},\downarrow}\left(-\frac{\partial f_{\mathbf{k}}^0}{\partial E_{\mathbf{k}}}\right) \bigg/ \sum_{\mathbf{k}}\left(-\frac{\partial f_{\mathbf{k}}^0}{\partial E_{\mathbf{k}}}\right). \quad \text{(S7)}$$

The spin relaxation in graphene due to local SOC for the Hamiltonian $V_s^{(2)}$, Eq. (S2), was analyzed in Ref. [S14]. In the present study we straightforwardly extend that calculation scheme to the spin-full particle-hole Nambu basis tailored to the superconducting graphene.

The case of magnetic impurity interacting via the exchange interaction with graphene electrons,

$$V_s^{(1)} = -J\mathbf{s}\cdot\mathbf{S} = -J\big(\sigma_x\otimes\Sigma_x + \sigma_y\otimes\Sigma_y + \sigma_z\otimes\Sigma_z\big), \quad \text{(S8)}$$

where $\mathbf{s} = (\sigma_x,\sigma_y,\sigma_z)$ acts in the spin space of electrons, and $\mathbf{S} = (\Sigma_x,\Sigma_y,\Sigma_z)$ in the spin space of impurity, is handled in details in the proceedings review [S12]. The spin-relaxation rate equals

$$\frac{1}{\tau_s} = \sum_{\mathbf{k},\mathbf{q}} W_{\mathbf{k},\uparrow,\Downarrow|\mathbf{q},\downarrow,\Uparrow}\left(-\frac{\partial f_{\mathbf{k}}^0}{\partial E_{\mathbf{k}}}\right) \bigg/ \sum_{\mathbf{k}}\left(-\frac{\partial f_{\mathbf{k}}^0}{\partial E_{\mathbf{k}}}\right), \quad \text{(S9)}$$

where $W_{\mathbf{k},\uparrow,\Downarrow|\mathbf{q},\downarrow,\Uparrow}$ is the probability per unit time for the spin-flip scattering process when an electron scatters from the state $\mathbf{q}$ to $\mathbf{k}$ and flips its spin from $\downarrow$ to $\uparrow$. Due to the conservation of angular momentum, the impurity flips its spin from $\Uparrow$ to $\Downarrow$. Discussions presented in Refs. [S12, S13] formulated the problem in the extended Hilbert space that counts the impurity spin degrees of freedom $\Sigma = \{\Uparrow,\Downarrow\}$, i.e., each original electron annihilation (creation) operator $c_{m,\sigma}^{(\dagger)}$ was extended by $\Sigma$ giving rise to $c_{m,\sigma,\Sigma}^{(\dagger)}$ acting in the extended space. Then, by making the unitary transformation in the spin space—$(\sigma,\Sigma)$ maps to the singlet and triplet states—one can diagonalize $V_s^{(1)}$, Eq. (S8), obtain T-matrices, and compute the transition rates. Finally, to get rid off the impurity spins, we traced out those rates by the impurity (spin-unpolarized) density matrix $\varrho = \frac{1}{2}|\Uparrow\rangle\langle\Uparrow| + \frac{1}{2}|\Downarrow\rangle\langle\Downarrow|$. The exactly same analytical procedure is carried out for the case of superconducting graphene. However, there is a shorter path to arrive at the same final results for *the spin-flip transition rates* that avoids the extension into enlarged Hilbert space and then tracing out its additional degrees of freedom. That is beneficial for the *Kwant implementation of the magnetic exchange problem*—the *same spin-flip transition rates* that would be generated by $V_s^{(1)}$ can be obtained from the simpler effective Hamiltonian

$$\tilde{V}_s^{(1)} = J\big(d_\uparrow^\dagger d_\uparrow + d_\downarrow^\dagger d_\downarrow\big) - 2J\big(d_\uparrow^\dagger d_\downarrow + d_\downarrow^\dagger d_\uparrow\big) \quad \text{(S10)}$$

that does not involve $\mathbf{S}$ degrees of freedom, just the annihilation (creation) operators $d_\sigma^{(\dagger)}$ acting on electrons at the adatom site in the same way as in Hamiltonian $V_o$ in the main text. As a comment, the form of $\tilde{V}_s^{(1)}$ is not unique and, as a warning, such an *effective* Hamiltonian $\tilde{V}_s^{(1)}$ would not reproduce the correct spin-conserving rates as the *original* exchange Hamiltonian $V_s^{(1)}$.

Our methodology for the *full analytical approach* as shortly outlined above, can be compactly and with all consequent logical steps presented as the following route map: Starting from $H_0 = H_g$ at given $\mu$ and $\Delta$ [see the Eq. (1) in main text], we compute:

**(1)** the associated unperturbed eigenspectrum of superconducting graphene $E_{\mathbf{k}} = \sqrt{(\epsilon_{\mathbf{k}}-\mu)^2 + \Delta^2}$, where $\epsilon_{\mathbf{k}}$ represents known normal-phase eigenvalues,

**(2)** the corresponding quasiparticle eigenstates $|\mathbf{k},\sigma\rangle$ serving as 'in' and 'out' scattering states [normalized as $\langle\mathbf{k},\sigma|\mathbf{q},\sigma'\rangle_{unit\ cell} = \delta_{\mathbf{k},\mathbf{q}}\delta_{\sigma,\sigma'}$],

**(3)** the unperturbed (retarded) Green's functions $\mathbb{G}_0 = (E - H_0 + i\epsilon)^{-1}$ consisting of normal and anomalous parts. From $\mathbb{G}_0$ and given $V = V_o + V_s^{(1)/(2)}$, we get T-matrix $\mathbb{T} = V(1 - \mathbb{G}_0 V)^{-1}$, as described above, which yields the scattering amplitudes $\langle\mathbf{k},\uparrow|\mathbb{T}|\mathbf{q},\downarrow\rangle$ and the perturbed Green's function $\mathbb{G} = \mathbb{G}_0 + \mathbb{G}_0\mathbb{T}\mathbb{G}_0$,

**(4)** from the full Green's function $\mathbb{G}$, we evaluate the (local) density of states (trace of $\mathbb{G}$), bound states (poles of $\mathbb{G}$), and other spectral characteristics of perturbed superconducting graphene,

**(5)** to access the spin relaxation rate $1/\tau_s^{\text{SC}}$ at given $\mu$, $T$, and $\eta$, we numerically evaluate [generalized Fermi's golden rule in the presence of thermal smearing]:

$$\frac{1}{\tau_s^{\text{SC}}} = \frac{\iint_{\text{BZ}} d\mathbf{k}\,d\mathbf{q}\,|\langle\mathbf{k},\uparrow|\mathbb{T}|\mathbf{q},\downarrow\rangle|^2\,\delta(E_{\mathbf{k}} - E_{\mathbf{q}})\left(-\frac{\partial g}{\partial E_{\mathbf{k}}}\right)}{\frac{\hbar\pi}{A_{uc}\eta}\int_{\text{BZ}} d\mathbf{k}\left(-\frac{\partial g}{\partial E_{\mathbf{k}}}\right)}, \quad \text{(S11)}$$

where the integrations are taken over graphene's 1st Brillouin zone (BZ); $g = [\exp(\frac{E_{\mathbf{k}}}{k_{\text{B}}T}) + 1]^{-1}$ is the Fermi-Dirac distribution (its derivative gives thermal smearing), while $A_{uc}$ is the area of graphene unit cell. Yafet's relation, Eq. (S1), emerges as a special case of Eq. (S11). Employing the first Born approximation

$$\mathbb{T} \simeq V_s = \sum_{\mathbf{k},\mathbf{q}} (v_s)_{\mathbf{kq}} c_{\mathbf{k}\uparrow}^\dagger c_{\mathbf{q}\downarrow} + \text{h.c.}, \quad \text{(S12)}$$

expressing the electron operators $c^\dagger$ and $c$ therein via the corresponding Bogoliubov QP operators $\gamma^\dagger$ and $\gamma$, and using $(v_s)_{-\mathbf{q}-\mathbf{k}} = \mp(v_s)_{\mathbf{kq}}$, which results from even/odd parity of $V_s$ w.r.t. time reversal, one gets

$$\langle\mathbf{k},\uparrow|\mathbb{T}|\mathbf{q},\downarrow\rangle \simeq (u_{\mathbf{k}}u_{\mathbf{q}} \mp v_{\mathbf{k}}v_{\mathbf{q}})(v_s)_{\mathbf{kq}}. \quad \text{(S13)}$$

Taking its absolute square, plugging it into Fermi's golden rule, and integrating over $\mathbf{q}$ and $\mathbf{k}$ (while accounting for the thermal smearing factor) yields the spin relaxation rate as given by Eq. (S1).

### Numerical *Kwant* implementation

For the numerical tight-binding calculations, we use the python package *Kwant* [S15]. In order to study the spin relaxation of QPs in the superconducting state, we employ the following procedure. The case of SOC perturbation is the more intricate one and will be addressed in detail. The original model Hamiltonian $H_g + V_o + V_s^{(2)}$, written in the local atomic basis, can be represented in the Bogoliubov-de Gennes form

$$H_{\text{BdG}} = \begin{pmatrix} H_G + V_s^{(2)} & i\sigma_y \Delta \\ -i\sigma_y \Delta^* & -[H_G + V_s^{(2)}]^* \end{pmatrix} \quad \text{(S14)}$$

$$= \begin{pmatrix} A & SO^{ee}_{\uparrow\downarrow} & 0 & \Delta \\ SO^{ee}_{\downarrow\uparrow} & B & -\Delta & 0 \\ 0 & -\Delta^* & -A^* & -(SO^{ee}_{\uparrow\downarrow})^* \\ \Delta^* & 0 & -(SO^{ee}_{\downarrow\uparrow})^* & -B^* \end{pmatrix}.$$

Its four blocks are given in the spin-full Nambu basis

$$\Psi = \left( \Psi^e_\uparrow,\ \Psi^e_\downarrow,\ \Psi^h_\uparrow,\ \Psi^h_\downarrow \right)^\top, \quad \text{(S15)}$$

where $\Psi^e_\sigma$ ($\Psi^h_\sigma$) stands, schematically, for an array of all annihilation (creation) operators with spin $\sigma$ that operate on local atomic orbitals residing on the real-space lattice (graphene + adatom). Components $A$, $B$, $SO^{ee}_{\uparrow\downarrow}$, and $SO^{ee}_{\downarrow\uparrow}$ represent spin-conserving and spin-flipping tight-binding matrix elements in the normal phase. Each of them is a matrix, whose entries are labeled by the two lattice sites' indices. For example, the element $A_{m,n}$ represents spin-conserving hopping of an electron with spin up from the site $n$ to site $m$. Since we are interested in the spin relaxation of QPs, a more convenient implementation in *Kwant* relies on the global unitary transformation

$$U = \begin{pmatrix} 1 & 0 & 0 & 0 \\ 0 & 0 & 0 & 1 \\ 0 & 0 & 1 & 0 \\ 0 & 1 & 0 & 0 \end{pmatrix}, \quad \text{(S16)}$$

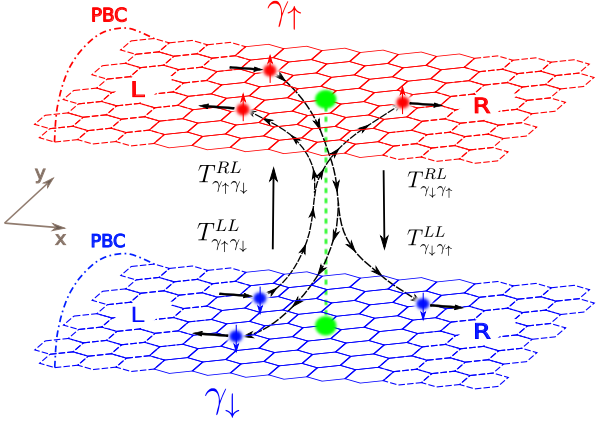

FIG. S2. *Kwant*-implementation scheme. Original graphene lattice including the adatom site is virtually split into two subsidiary QP lattices, upper (red color) for spin-up QPs $\gamma_\uparrow$ and lower (blue color) for spin-down QPs $\gamma_\downarrow$—their dynamics are governed by diagonal parts of Eq. (S19). The green areas denote the impurity site, namely its spin-up and spin-down degrees of freedom embedded within the corresponding QP-up and QP-down layers. The spin-flip part of the perturbation $V_s$ is local in space and couples those two virtual QP layers—this is our local scattering region and its spin-flip dynamics is dictated by the off-diagonal part of Eq. (S19). We assume periodic boundary conditions (PBCs) along the transverse $y$-direction—such quasi-1D geometry with PBCs effectively simulates 2D transport [S14]. Left (L) and right (R) regions in the QP-up and QP-down virtual layers represent leads that are injecting and extracting QPs with the corresponding spin degree of freedom and the required propagation direction. For each incoming mode—entering from L or R into the QP-up or QP-down layer—we extract the corresponding spin-flip transmission amplitudes—$T^{RL}_{\gamma_\uparrow \gamma_\downarrow}$, $T^{RL}_{\gamma_\downarrow \gamma_\uparrow}$, $T^{LR}_{\gamma_\uparrow \gamma_\downarrow}$, $T^{LR}_{\gamma_\downarrow \gamma_\uparrow}$, $T^{LL}_{\gamma_\uparrow \gamma_\downarrow}$, $T^{LL}_{\gamma_\downarrow \gamma_\uparrow}$, $T^{RR}_{\gamma_\uparrow \gamma_\downarrow}$, and $T^{RR}_{\gamma_\downarrow \gamma_\uparrow}$—components of the S-matrix as provided by *Kwant*. As a great advantage, this virtual-layer implementation with well-defined QP spins does not involve reflection amplitudes. Two scattering processes (with L-incoming and L-/R-outgoing states, together with their relevant $T$-amplitudes) are for the sake of visualization displayed by the black dashed paths—spin flips inevitably displace QPs between the two virtual layers. Having the spin-flip components of the S-matrix, or equivalently those of the T-matrix, at given QP energy $E$, we obtain the average spin-flip probability $p_s(E)$ from Eq. (S24).

which transforms the Hamiltonian $H_{\text{BdG}}$ into

$$H'_{\text{BdG}} = \begin{pmatrix} A & \Delta & 0 & SO^{ee}_{\uparrow\downarrow} \\ \Delta^* & -B^* & -(SO^{ee}_{\downarrow\uparrow})^* & 0 \\ 0 & -(SO^{ee}_{\uparrow\downarrow})^* & -A^* & -\Delta^* \\ SO^{ee}_{\downarrow\uparrow} & 0 & -\Delta & B \end{pmatrix}. \quad \text{(S17)}$$

This representation is tailored to the basis $(\gamma_\uparrow, \gamma_\downarrow)^\top$, consisting of two reduced particle-hole Nambu spinors

$$\gamma_\uparrow = \left( \Psi^e_\uparrow,\ \Psi^h_\downarrow \right)^\top \quad \text{and} \quad \gamma_\downarrow = \left( \Psi^h_\uparrow,\ \Psi^e_\downarrow \right)^\top. \quad \text{(S18)}$$

The above transformation eventually partitions the Hamiltonian into two blocks—the up-spin QP sector $\gamma_\uparrow$ (*upper-left* corner) and the down-spin QP sector $\gamma_\downarrow$ (*lower-right* corner). Those blocks are coupled by the off-diagonal spin-flip perturbation Hamiltonian (sparse matrix) that encompasses just the spin-flip part of the local SOC (or magnetic exchange). The total Hamiltonian takes therefore the form

$$H'_{\text{BdG}} = \begin{pmatrix} H_{\gamma_\uparrow \gamma_\uparrow} & H^{SO}_{\gamma_\uparrow \gamma_\downarrow} \\ H^{SO}_{\gamma_\downarrow \gamma_\uparrow} & H_{\gamma_\downarrow \gamma_\downarrow} \end{pmatrix}. \quad \text{(S19)}$$

Making use of $H'_{\text{BdG}}$ significantly simplifies not only the implementation of the QP scattering problem

in *Kwant* [see also Fig. S2 and the detailed description in its caption], but also the interpretation of the obtained results in terms of the S-matrix entries. The reduced Nambu basis naturally separates the QP spins, and, more importantly, allows us to directly compute the superconducting S-matrix, which is not so obvious in the original basis due to *Kwant* code specific technicalities ("conservation laws").

In order to effectively simulate 2D bulk transport employing the 1D ribbon calculation, we follow the method presented, for example, in Ref. [S14]. Particularly, we impose $\Phi$-periodic boundary conditions (PBCs) along the transverse $y$-direction that promote hoppings between the sites $(x, y = 0)$ and $(x, y = W)$ lying on opposite edges of the 1D ribbon ($W$ stands for its width). Those hoppings enter $H_{\gamma_\uparrow \gamma_\uparrow}$ and $H_{\gamma_\downarrow \gamma_\downarrow}$ terms in Eq. (S19). For the sake of brevity, we represent them in the tight-binding form using $\gamma_\uparrow$ and $\gamma_\downarrow$ operators at sites $(x, 0)$ and $(x, W)$:

$$\text{hop}^{\Phi\text{-PBC}} = \gamma^\dagger_{\uparrow,(x,W)} \begin{pmatrix} -te^{-i\Phi} & 0 \\ 0 & te^{-i\Phi} \end{pmatrix} \gamma_{\uparrow,(x,0)}$$
$$+ \gamma^\dagger_{\downarrow,(x,W)} \begin{pmatrix} te^{+i\Phi} & 0 \\ 0 & -te^{+i\Phi} \end{pmatrix} \gamma_{\downarrow,(x,0)} + \text{h.c.}, \quad \text{(S20)}$$

where the angle $\Phi \in [0, 2\pi]$, and $t$ is the nearest neighbor graphene hopping. Note that the phase factors in the $\gamma_\downarrow$-part should be complex conjugated in order to preserve time-reversal symmetry of the unperturbed superconducting host. This additional phase factors allow for accessing different transverse momenta $k_y$ and thus for the simulation of 2D transport characteristics. A detailed scheme how that works is illustrated in Fig. S3 for ordinary graphene. Finite width of the scattering region discretizes the allowed values of the transverse momenta $k_y$ and the resulting 1D band structure foliates into several projected subbands. Changing $\Phi$ in the PBCs, the originally discrete $k_y$ levels in **k**-space shift by an equal amount and different sets of 2D graphene states can therefore be addressed. Averaging over many different phases basically leads to a reliable sampling of 2D graphene's Brillouin zone and hence to bulk transport simulations which can be well compared with analytical results. The aforementioned strategy and implementation is used to numerically obtain the S-matrix elements.

To treat the magnetic moment-mediated spin relaxation of QPs, we proceed in exactly the same way, i.e., we start from the Hamiltonian $H_{\text{BdG}}$, Eq. (S14), but we use the effective Hamiltonian $\tilde{V}_s^{(1)}$, Eq. (S10), instead of $V_s^{(2)}$, and then follow the same steps as in the above SOC case.

### Spin relaxation rates from *Kwant*

Knowing the S-matrix entries as functions of chemical potential, the QP energy, and the incoming and outgoing modes, we can compute the QP spin-flip rate $1/\tau_s(E)$ at energy $E$, and consequently the total spin-flip rate $1/\tau_s(\mu, T)$ in the superconducting phase held at chemical potential $\mu$ and temperature $T$. The fermionic nature of QP excitations implies that those two rates are related [S2, S16] via the formula

$$\frac{1}{\tau_s(\mu, T)} = \frac{\int\limits_\Delta^\infty dE\, \text{QP-DOS}_{sr}(E) \left(-\frac{\partial g}{\partial E}\right) \frac{1}{\tau_s(E)}}{\int\limits_\Delta^\infty dE\, \text{QP-DOS}_{sr}(E) \left(-\frac{\partial g}{\partial E}\right)}, \quad \text{(S21)}$$

where the integration is taken over all QP energies, QP-DOS$_{sr}(E)$ represents the unperturbed QP density of states in the *scattering region* at energy $E$, and $g(E, T) = [\exp(E/k_\text{B} T) + 1]^{-1}$ is the Fermi-Dirac distribution [note that QP energies are measured relative to $\mu$ since the latter explicitly enters the model Hamiltonian $H_g$]. In what follows, we derive the expressions for $1/\tau_s(E)$ and QP-DOS$_{sr}(E)$ employing solely the entries accessible from the *Kwant* package, and then, using Eq. (S21), we compare the numerical approach with our analytical calculations.

*Kwant* provides S-matrix entries among different leads' modes that are labeled by spin $\sigma$, energy $E$, and a set $\{n\}$ of other discrete indices. For example, in our 1D problem the latter represent the transverse lead mode quantum number and the direction of motion—encoded in the superscripts L and R [convention used in Fig. S2: incoming (outgoing) L-/R-mode departs from (arrives at) L-/R-lead]. In sequel, we suppress most of them and, for the sake of brevity, we use just collective labels $i$ and $o$ for the incoming and outgoing modes, and the spin index $\sigma = \{\uparrow, \downarrow\}$. Our system, displayed in Fig. S2, consists of spin-up and spin-down QP semi-infinite leads with width $W$, and a scattering region with length $L$ and width $W$ that hosts $4WL/(\sqrt{3}a^2)$ carbon atoms [graphene lattice constant $a = 2.46\,\text{Å}$] and one impurity, hence concentration

$$\eta = \frac{\sqrt{3}a^2}{4WL}. \quad \text{(S22)}$$

Assuming that there is an *in*coming QP mode $(i, \sigma)$ at energy $E$, the probability $p_{-\sigma,\sigma}(i, E)$ measuring that the mode $(i, \sigma)$ experiences a spin-flip equals

$$p_{-\sigma,\sigma}(i, E) = \sum_o \left|S^{o,i}_{-\sigma,\sigma}(E)\right|^2, \quad \text{(S23)}$$

where the sum over $o$ runs over all *out*going modes $(o, -\sigma)$ at energy $E$. Then, the average QP spin-flip probability at energy $E$ reads

$$p_s(E) = \frac{1}{2\#_i} \sum_i p_{\uparrow,\downarrow}(i, E) + \frac{1}{2\#_i} \sum_i p_{\downarrow,\uparrow}(i, E), \quad \text{(S24)}$$

where the term in the denominator $\#_i$ counts the number of accessible incoming modes per spin at energy $E$. The

factor 1/2 takes into account the probability—picking a QP at energy $E$, it is equally likely to have up- or down-spin.

It is specific to the *Kwant* package to associate with each incoming mode $(i, \sigma)$ the lead wave function $|\Psi_{i,\sigma}\rangle$, such that its scalar product over the lead-unit cell [that covers spatial area $a \times W$ and the number of $2W/(\sqrt{3}a)$ of graphene unit cells along the transverse direction] is given as follows

$$\langle \Psi_{i,\sigma} | \Psi_{i,\sigma} \rangle_{luc} = \frac{1}{\hbar} \frac{a}{v_g(i)}, \quad \text{(S25)}$$

where $v_g(i)$ stands for the magnitude of group velocity of the mode $(i, \sigma)$. Since the underlying system is not magnetic, the corresponding QP energies and group velocities are same for both spin projections. The *Kwant* normalization of $|\Psi_{i,\sigma}\rangle$ implies that the QP DOS in the lead unit cell equals

$$\text{QP-DOS}_{luc}(E) = \frac{1}{2\pi} \sum_{(i,\sigma)} \langle \Psi_{i,\sigma} | \Psi_{i,\sigma} \rangle_{luc}, \quad \text{(S26)}$$

and, therefore, the unperturbed QP DOS in the scattering region $L \times W$ is given as follows

$$\text{QP-DOS}_{sr}(E) = \left(\frac{L}{a}\right) \text{QP DOS}_{luc}(E). \quad \text{(S27)}$$

The average time $t(E)$, the incoming QP at energy $E$ would need in order to scatter into a family of empty outgoing modes would be the average of $L/v_g(i)$ over all incoming modes, i.e.,

$$t(E) = \frac{1}{2 \#_i} \sum_{(i,\sigma)} \frac{L}{v_g(i)} = \frac{L/a}{2 \#_i} \sum_{(i,\sigma)} \frac{a}{v_g(i)} \quad \text{(S28)}$$
$$= \hbar \frac{L/a}{2 \#_i} \sum_{(i,\sigma)} \langle \Psi_{i,\sigma} | \Psi_{i,\sigma} \rangle_{luc} = \frac{2\pi\hbar}{2 \#_i} \text{QP-DOS}_{sr}(E),$$

where we have used Eqs. (S25)-(S27). Dividing the average spin-flip probability $p_s(E)$ by the average time $t(E)$ gives the average QP spin-flip rate at energy $E$,

$$\frac{1}{\tau_s(E)} = \frac{1}{2\pi\hbar} \frac{\sum_{i,o} |S^{o,i}_{\uparrow,\downarrow}(E)|^2 + |S^{o,i}_{\downarrow,\uparrow}(E)|^2}{\text{QP-DOS}_{sr}(E)}. \quad \text{(S29)}$$

Plugging the partial results, Eqs. (S26) (S27) and (S29), into the main formula for the spin-relaxation rate $1/\tau_s(\mu, T)$, Eq. (S21), we get the central result of this section—the Kochan-Barth formula:

$$\frac{1}{\tau_s^{Kw}} = \frac{\frac{a}{L}}{\hbar} \frac{\int_\Delta^\infty dE \left(-\frac{\partial g}{\partial E}\right) \sum_{i,o,\sigma} |S^{o,i}_{-\sigma,\sigma}(E)|^2}{\int_\Delta^\infty dE \left(-\frac{\partial g}{\partial E}\right) \sum_{(i,\sigma)} \langle \Psi_{i,\sigma} | \Psi_{i,\sigma} \rangle_{luc}}. \quad \text{(S30)}$$

The quantities on the right-hand side, the entries of S-matrix, and the lead wave functions are fully accessible from the *Kwant* calculation. Multiplying the nominator of the above fraction by unity, written in the form $(4W)/(\sqrt{3}a) \times (\sqrt{3}a)/(4W)$, we can write

$$\frac{1}{\tau_s^{Kw}} = \frac{\eta \frac{4W}{\sqrt{3}a}}{\hbar} \frac{\int_\Delta^\infty dE \left(-\frac{\partial g}{\partial E}\right) \sum_{i,o,\sigma} |S^{o,i}_{-\sigma,\sigma}(E)|^2}{\int_\Delta^\infty dE \sum_{(i,\sigma)} \langle \Psi_{i,\sigma} | \Psi_{i,\sigma} \rangle_{luc} \left(-\frac{\partial g}{\partial E}\right)}. \quad \text{(S31)}$$

The energy integration is numerically evaluated employing the Riemann summation, where we integrate from the band gap $\Delta$ to a energy cut-off of $\Delta + 25 k_B T$. In order to have a higher resolution close to the band gap $\Delta$, we split the integral into two integration regions. First, we integrate from $\Delta$ to $\Delta + 3.5 k_B T$ with 15 steps and second, to $\Delta + 25 k_B T$ with 25 steps. We keep the same partitioning prescription for all temperatures keeping the same number of points within the corresponding regions, so that all integrated results have the same "systematic error". To keep a reasonable balance between the computational efforts and the required accuracy, we restrict our calculations to a scattering region of length $L = 5\,a$ and width $W = 299\,a$ (we checked that without including a disorder the dependence on $L$ is substantially less pronounced than that on $W$, there we use such elongated geometry). For averaging over the angle that enters $\Phi$-PBC, we choose 20 equally distributed values within $[0, 2\pi]$. Given our parameters, the corresponding $\eta_{Kwant}$ equals 288 ppm.

The results of the *Kwant* package calculations are presented in Figs. S4 and S5, respectively, along with the corresponding data from the analytic T-matrix approach presented in the opening subsection to that paragraph. Similarly as in the main text, Fig. S4 displays the magnetic-impurity-scenarios and Fig. S5 the SOC-dominated regimes, each for hydrogen [panels (a)] and fluorine [panels (b)]. Comparing the *Kwant*-computed results with the analytical ones, as also presented in the main text [see Figs. 2(a) and (b), as well as Fig. 3 therein], we see the striking qualitative and quantitative agreements between both approaches over the whole ranges of doping and temperature. All distinguished characteristics as observed analytically occur simultaneously in the numerical *Kwant* simulations, serving as an unprecedented evidence of our predictions' self-consistency. Therefore, we believe that our findings are universal and potentially interesting for spin-relaxation experiments that can, then, resolve the dominant spin-relaxation mechanism in graphene.

As a comment, the numerically determined spin-relaxation curves are not as smooth as the analytical ones since we need to restrict ourselves to reasonable scattering region sizes; otherwise, the computational efforts would become unmanageable and too expensive.

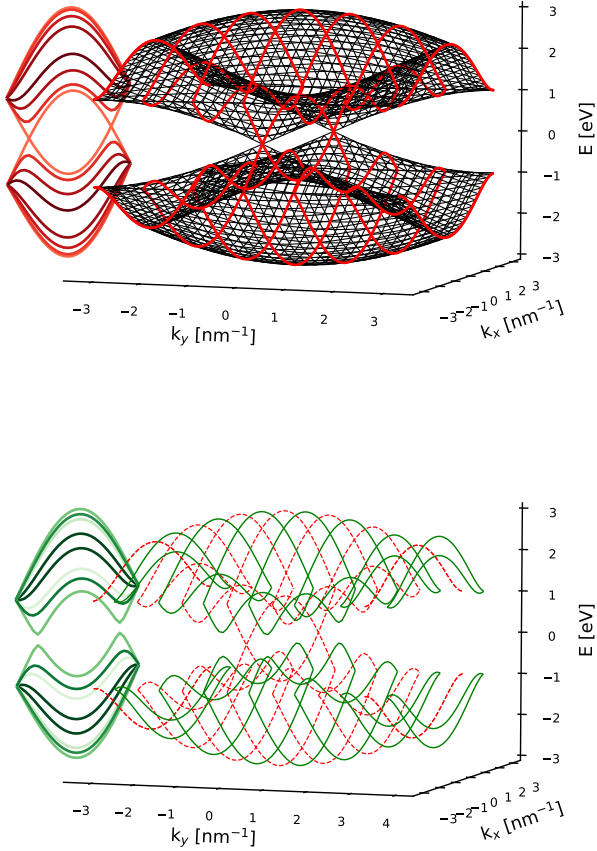

FIG. S3. The phase-averaging scheme with $\Phi$-periodic boundary conditions for graphene in the normal phase. *Top:* Electronic band structure of 2D graphene (3D black mesh) versus momenta $(k_x, k_y)$. The finite width of the sample along the $y$-direction quantizes the allowed values of $k_y$ what is giving the discretized electronic dispersion (red lines on top of the 3D black mesh). Projecting the latter along $k_y$ in the Brillouin zone gives the distinct $k_x$-dependent projected subbands shown on the left (displayed in different tones of red). *Bottom:* Changing the phase of $\exp(\pm i\Phi)$ in the PBCs shifts the discretized electronic dispersion—schematically, the original red bands move to the green ones. The resulting subbands give rise to the new projected band structure (displayed on the left in different tones of green). Sampling $\Phi \in [0, 2\pi]$ covers different incoming and outgoing modes by allowing different transverse momenta $k_y$. This enables us to simulate 2D bulk transport characteristics within a 1D ribbon geometry.

The limited size of the scattering region and the aforementioned phase averaging over the transverse modes raise some numerical fluctuations, which—although being clearly evident—do not spoil the qualitative behavior. Moreover, finite-size effects cause that the value of $\eta$ needed to align analytical calculations (in most cases 280 ppm) with the numerical results slightly deviates from the corresponding $\eta_{Kwant}$ (288 ppm). The finite size effects are mostly visible in the case of SOC in hydrogenated graphene, Fig. S5(a). This is because the typical energy separation due to finite width $W$ is about 5 meV while the dominant SOC scale $\Lambda_{\mathrm{PIA}}^{\mathrm{B}} \simeq 1$ meV, in that case to match numeric with analytic we need effectively lower $\eta = 220$ ppm.

## (*ANTI-*)HEBEL-SLICHTER EFFECT IN RESONANT MAGNETIC JOSEPHSON JUNCTIONS

In the main text, we attributed the breakdown of the Hebel-Slichter effect in superconducting graphene to a substantial energy separation between the QP excitations that reside at the coherence peaks and YSR bound states around impurities whose energies immerse deep inside the superconducting gap. It is well known (see, for example Ref. [S17] and references therein) that such YSR states determine underlying spectral and spin characteristics. For instance, they are responsible for 0-$\pi$ transitions in magnetic Josephson junctions that reverse the direction of the supercurrent. This section demonstrates yet on another example—*resonant magnetic* Josephson junctions—that the observed breakdown of the Hebel-Slichter effect seems to be a *universal physical phenomenon* that accompanies QP spin transport through magnetic resonances in $s$-wave superconductors.

We consider the *one-dimensional* Josephson geometry depicted in Fig. S6(a). Two semi-infinite superconductors are coupled via the "*resonant* superconducting trap" with width $a$ (not related to the graphene lattice constant). The latter is composed of two outer ultrathin deltalike scalar tunnel barriers—residing at $z = 0$ and $z = a$, respectively—and the additional host—deltalike magnetic impurities—intercalated in the middle, i.e., at $z = a/2$. Such a configuration enables a full analytical treatment, yielding a relatively simple and visual toy model of *resonant magnetic impurity scattering* inside Josephson junctions that allows to qualitatively and quantitatively explore its spin-flip properties.

Scattering of QPs and the formation of YSR states in the above resonant setup can be described in terms of the Bogoliubov-de Gennes Hamiltonian [S16] that needs to be formulated within the spin-full Nambu basis,

$$H_{\mathrm{BdG}} = \begin{bmatrix} H_e & \Delta(z) \\ \Delta^\dagger(z) & H_h \end{bmatrix}. \quad (S32)$$

It couples the single electronlike QP Hamiltonian

$$H_e = \left(-\frac{\hbar^2}{2m}\frac{\mathrm{d}^2}{\mathrm{d}z^2} - \mu\right)\sigma_0 + \lambda_{\mathrm{SC}}\sigma_0\big[\delta(z) + \delta(z-a)\big] + \big[\lambda_{\mathrm{MA},z}\sigma_z + \lambda_{\mathrm{MA},x}\sigma_x\big]\delta(z-a/2) \quad (S33)$$

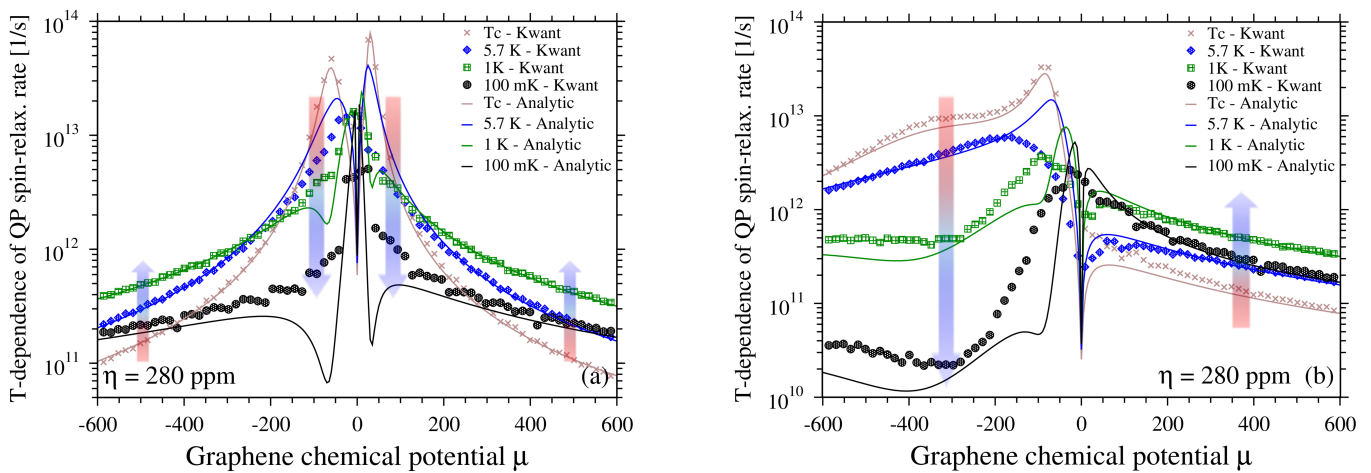

FIG. S4. Comparison of *Kwant* (symbols) and analytical (solid lines) calculations of QP spin-relaxation rates in superconducting graphene in the presence of hydrogen [panel (a)] and fluorine [panel (b)] magnetic impurities with concentration (per carbon atom) $\eta_{Kwant} = 288$ ppm. Both methods match and provide consistent qualitative and quantitative pictures regarding the temperature and doping dependencies of $1/\tau_s$, and its enhancement and depletion depending on the position of the Fermi level $\mu$ with respect to resonances. Rainbow arrows indicate the expected increase of the spin relaxation with lowered temperature outside the resonances—the Hebel-Slichter effect—and its predicted break-down [substantial decrease of the spin-relaxation] near the resonances due to emergent YRS states deep inside the superconducting gap. Magnetic exchange coupling is $J = -0.4$ eV for hydrogen, and $J = 0.5$ eV for fluorine, the *Kwant* simulation used a scattering region with $W = 299\,a$ and $L = 5\,a$. Analytical results require slightly lower concentrations $\eta$ [values are inside the plots] to be perfectly aligned with the *Kwant* data. The reasons for that are originating from finite-size effects and the finite number of phases used in the PBC averaging.

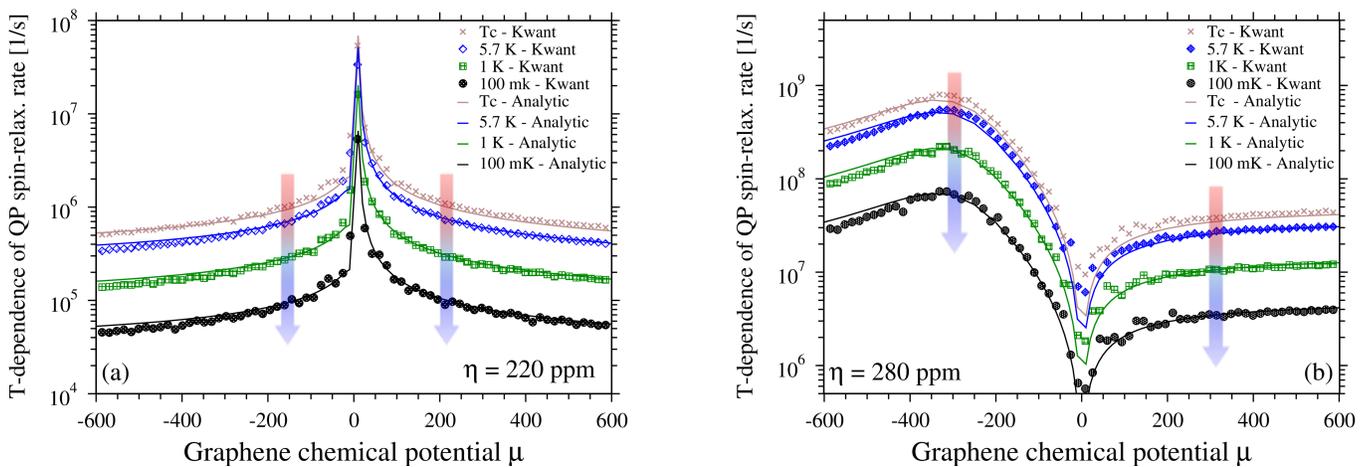

FIG. S5. Comparison of *Kwant* (symbols) and analytical (solid lines) calculations of QP spin-relaxation rates in superconducting graphene in the presence of locally enhanced SOC for hydrogen [panel (a)] and fluorine [panel (b)] impurities with $\eta_{Kwant} = 288$ ppm. Both methods match perfectly and provide the same qualitative and quantitative features regarding the temperature and doping dependencies of $1/\tau_s$. Rainbow arrows evolving from red to blue indicate the expected decreasing trend of the spin relaxation with lowered temperature. SOC parameters are given in Table S1 and the geometrical setup is the same as in Fig. S4. Analytical results require slightly lower concentrations $\eta$ [values are inside the plots] to be perfectly aligned with the *Kwant* data [$W = 299\,a$ and $L = 5\,a$]. The reasons for that are originating from finite-size effects and the finite number of phases used in the PBC averaging.

with its holelike counterpart $H_h = -\sigma_y H_e^* \sigma_y$, enwrought through the $s$-wave superconducting pairing potential $\Delta(z) = \Delta\,\sigma_0$. We assume the outer (semi-infinite) superconductors possess the *same* superconducting energy gap $\Delta$ and that the junction is not subject to any superconducting phase difference. To further simplify the analytical treatment, we consider also equal QP masses $m$ and chemical potentials $\mu$. The scattering strengths of the two identical outer barriers are parameterized by the scalar $\lambda_{\rm SC}$, whereas the magnetic scattering strengths in the middle are governed by $\lambda_{{\rm MA},z}$ and $\lambda_{{\rm MA},x}$, accordingly. The latter acts perpendicular

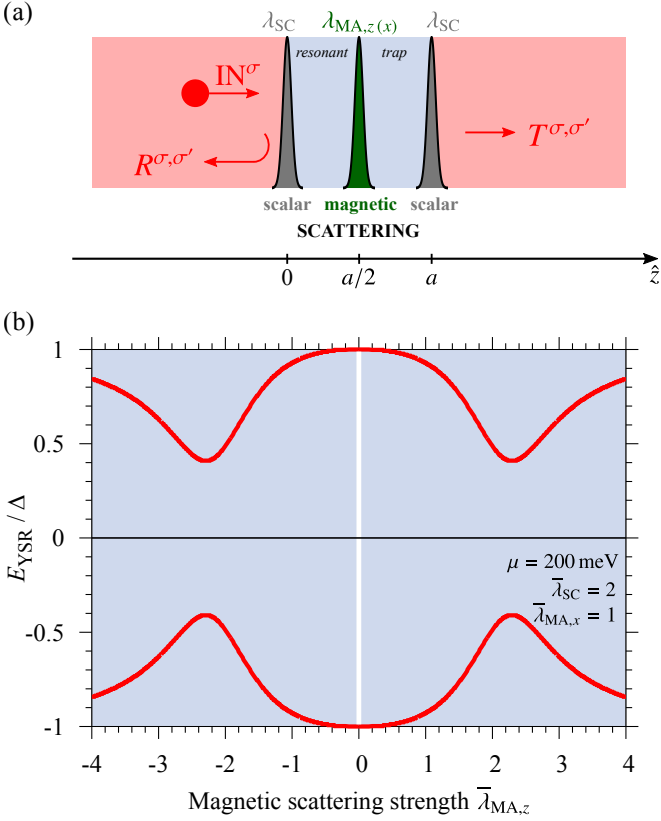

FIG. S6. (a) Schematic sketch of the *one-dimensional resonant magnetic Josephson junction* (oriented along $\hat{z}$), coupling two semi-infinite superconducting electrodes (light red) via the *resonant* superconducting trap (light blue) with width $a$. The latter consists of two outer deltalike tunneling barriers at $z = 0$ and $z = a$, parameterized by scalar strength $\lambda_{\text{SC}}$, and the central deltalike barrier at $z = a/2$ mimicking magnetic impurity scattering, parameterized by $\lambda_{\text{MA},z}$ and $\lambda_{\text{MA},x}$, respectively. An Incident QP with spin $\sigma$—IN$^\sigma$—arrives from the left superconductor and can be reflected back as a QP with spin $\sigma'$—probability $R^{\sigma,\sigma'}$—or transmitted through the resonant trap—probability $T^{\sigma,\sigma'}$. (b) Calculated energies of the YSR bound states, $E_{\text{YSR}}$, normalized to the magnitude $\Delta$ of the superconducting gap, as a function of the (dimensionless) magnetic strength $\overline{\lambda}_{\text{MA},z} = (2m\lambda_{\text{MA},z})/(\hbar^2 q_\text{F})$. We assume fixed scalar and magnetic $\lambda_{\text{MA},x}$ scattering strengths such that $\overline{\lambda}_{\text{SC}} = (2m\lambda_{\text{SC}})/(\hbar^2 q_\text{F}) = 2$ and $\overline{\lambda}_{\text{MA},x} = (2m\lambda_{\text{MA},x})/(\hbar^2 q_\text{F}) = 1$, as well the representative chemical potential $\mu = 200\,\text{meV} = 200\Delta_0 \gg \Delta_0$. The width of the resonant trap is $a = 10\,\text{nm}$. The magnetic impurity strength controls a wide parameter region in which the YSR states are *immersed deep inside the superconducting gap* (emphasized by shading).

to the QPs initial spin quantization axis ($\hat{z}$) and induces thus spin-flip scattering. As usual, $\sigma_0$ denotes the two-by-two identity matrix and $\sigma_i$ the $i$th Pauli matrix, all acting in spin space.

To unravel the energies of the YSR states that form around the central magnetic barrier of the junction, we follow the same strategy as described in Ref. [S17]. Regarding the model parameters, we use the same superconducting gap and the critical temperature as for superconducting graphene, i.e., $\Delta_0 = 1\,\text{meV}$ and $T_c \simeq 6.593\,\text{K}$. Furthermore, we take a representative chemical potential, $\mu = 200\,\text{meV} = 200\Delta_0 \gg \Delta_0$, with the corresponding Fermi wave vector $q_\text{F} = \sqrt{2m\mu}/\hbar$. For the scalar and magnetic $\lambda_{\text{MA},x}$ scattering strengths, we impose the moderate values $\overline{\lambda}_{\text{SC}} = (2m\lambda_{\text{SC}})/(\hbar^2 q_\text{F}) = 2$ and $\overline{\lambda}_{\text{MA},x} = (2m\lambda_{\text{MA},x})/(\hbar^2 q_\text{F}) = 1$—as motivated by [S18]—, while the resonant trap's width is set to $a = 10\,\text{nm}$. Figure S6(b) displays the computed energies of the subgap states as a function of the magnetic scattering strength $\overline{\lambda}_{\text{MA},z} = (2m\lambda_{\text{MA},z})/(\hbar^2 q_\text{F})$, confirming the previously observed behavior [S17]. As long as $\overline{\lambda}_{\text{MA},z}$ inside the trap is smaller than $|\overline{\lambda}_{\text{MA},z}| \lesssim 0.04$, the bound state spectrum contains only the usual Andreev states at $E_{\text{ABS}} = \pm\Delta$ and lacks the YSR counterparts. Contrary, going over that value, $|\overline{\lambda}_{\text{MA},z}| \gtrsim 0.04$, the spectrum starts to host also *YSR states*. Their energies are deeply immersed inside the gap so that one can expect reduced spin-flip scattering. Magnetic Josephson junctions' spectral properties are effectively tuned by changing $\overline{\lambda}_{\text{MA},z}$, which serves as the control parameter for tuning the energies of the YSR states in this section. Contrary to the case of graphene, where that was done by varying the chemical potential.

In the main text, we argued that magnetic impurities are expected to significantly impact the *QP spin-flip scattering characteristics* in resonances, and thereby cause the breakdown of the Hebel-Slichter effect. That this is not a red herring inherent to the nature of superconducting graphene is demonstrated now by the 1D resonant magnetic Josephson junction with $|\overline{\lambda}_{\text{MA},z}| \gtrsim 0.04$, supporting deeply immersed YRS states. Instead of computing the QP spin-relaxation rate $1/\tau_s$, which can be measured directly in experiments, we compute a from the theoretical viewpoint equivalent physical quantity, namely the *QP spin-flip scattering probability* $P_{\text{spin-flip}}$. The latter is given by

$$P_{\text{spin-flip}} = \sum_\sigma \sum_{\sigma' = -\sigma} \frac{1}{2}$$

$$\times \frac{\int_\Delta^{E'} \mathrm{d}E \left[R^{\sigma,\sigma'}(E) + T^{\sigma,\sigma'}(E)\right] \text{QP DOS}(E) \left(-\frac{\partial g}{\partial E}\right)}{\int_\Delta^{E'} \mathrm{d}E \, \text{QP DOS}(E) \left(-\frac{\partial g}{\partial E}\right)},$$

(S34)

counting the same spin-flip scattering processes that would determine the spin-relaxation rate, averaged over all excited QP states starting from the ground state energy $\Delta$. Regarding the computational details, we consider an incoming QP with spin $\sigma$ that enters from the left superconducting electrode and can either be reflected back or transmitted through the resonant magnetic trap into the right superconductor as a QP with

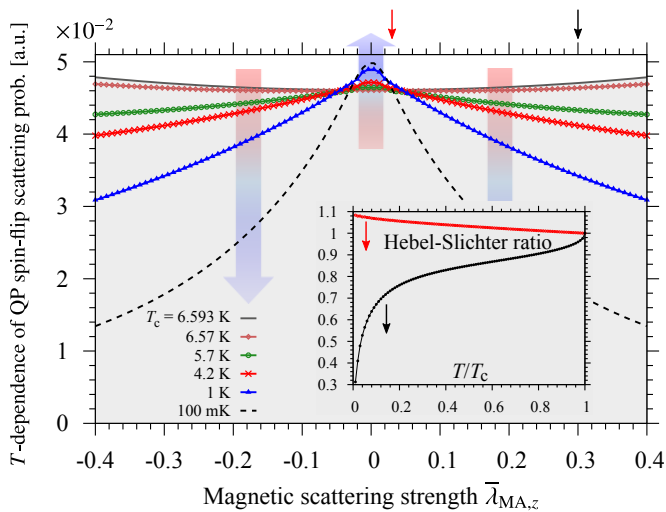

FIG. S7. QP spin-flip scattering probability at different temperatures and as a function of $\bar{\lambda}_{\text{MA},z}$; the remaining parameters are the same as in Fig. S6(b). With decreasing temperature, the spin-flip scattering probabilities *increase in the region where the YSR states overlap with the edges of the superconducting gap* and *decrease in the regions where the YSR states move into the gap*. The shaded region shows the related normal-phase spin-flip probability (as a guide for eyes). Rainbow arrows (coded in colors of temperature descent) indicate the raising (upward arrow) and lowering (downward arrow) trends of the spin-flip scattering probability as temperature goes down, starting at $T_c$ (normal phase). The inset shows the corresponding Hebel-Slichter ratio, $P_{\text{spin-flip}}(T)/P_{\text{spin-flip}}(T_c)$, as a function of $T/T_c$ and at two representative magnetic scattering strengths (indicated by red and black arrow ticks on the top horizontal axis): $\bar{\lambda}_{\text{MA},z} = 0.03$ (see red circled data) and $\bar{\lambda}_{\text{MA},z} = 0.3$ (see black circled data).

spin $\sigma' = -\sigma$. The scattering probabilities for the reflection and transmission of the incident QP with energy $E$ are denoted by $R^{\sigma,\sigma'}(E)$ and $T^{\sigma,\sigma'}(E)$, respectively. They are computed in the standard way as absolute squares of the amplitudes that come from the standard wave-function matching at each delta barrier. Recall that the QP DOS is given by QP DOS$(E) = \frac{E}{\sqrt{E^2-\Delta^2}} \text{DOS}(\mu)$ and the temperature dependence of the superconducting gap scales according to the BCS theory [S16], i.e., $\Delta = \Delta_0 \tanh(1.74\sqrt{T_c/T - 1})$. The upper limits in the above integrals should be infinities, but because of the thermal smearing $(-\frac{\partial g}{\partial E})$, the dominant contributions come from those QP states that reside in the coherence peaks. It is therefore most important to cover primarily that energy range, recognizing that the coherence peaks' energy widths scale $\sim \Delta$. Since the chemical potential $\mu$ is set to a large value anyway, we take $E' = \sqrt{\mu^2 + \Delta^2} \gg \Delta$ as the upper integration limits in the above formula for $P_{\text{spin-flip}}$.

Figure S7 presents the calculated QP spin-flip scattering probabilities as functions of temperature and the magnetic scattering strength $\bar{\lambda}_{\text{MA},z}$. We decrease the overall temperature in the same way as for superconducting graphene, see Figs. 2 and 3 in the main text. The range of $\bar{\lambda}_{\text{MA},z}$ is rather narrow to cover the crossover from the usual Hebel-Slichter scenario to its resonant breakdown. Comparing the outcomes with the YSR energies in Fig. S6(b), we observe the same trends as initially predicted for superconducting graphene. Hence, we can state the following conclusions: (1) superconducting systems hosting magnetic impurities support the *usual Hebel-Slichter effect* only when the YSR states overlap or are very close to the QP coherence peaks (for the considered junction, $|\bar{\lambda}_{\text{MA},z}| \lesssim 0.04$). In these cases, the related QP spin-flip scattering probabilities increase with decreasing temperature. Contrary to that, (2) once the strength of $\bar{\lambda}_{\text{MA},z}$ (or another control parameter) starts to immerse the *energies of YSR state deep inside the superconducting gap* (for the considered junction, $|\bar{\lambda}_{\text{MA},z}| \gtrsim 0.04$), *the Hebel-Slichter effect breaks down* and lowered temperature depletes the QP spin-flip processes.

Our findings serve as yet further evidence that supports the physical explanation of the Hebel-Slichter effect's breakdown at magnetic resonances and gives our findings more universal validity not limited just to graphene. Investigating the temperature dependence of QP spin relaxation in superconducting systems might provide a novel experimental tool to unravel the dominant spin-relaxation mechanisms and simultaneously sheds new light on the physics of YSR states.

---



\end{document}